# 1 Billion Pages = 1 Million Dollars?
# Mining the Web to Play "Who Wants to be a Millionaire?"

**Shyong (Tony) K. Lam**
Computer Science Dept.
University of Minnesota
Minneapolis, MN 55455
lam@cs.umn.edu

**David M. Pennock***
Overture Services, Inc.
74 N. Pasadena Ave., 3rd floor
Pasadena, CA 91101
david.pennock@overture.com

**Dan Cosley**
Computer Science Dept.
University of Minnesota
Minneapolis, MN 55455
cosley@cs.umn.edu

**Steve Lawrence**
NEC Laboratories America
Princeton, NJ 08540
lawrence@necmail.com

## Abstract

We exploit the redundancy and volume of information on the web to build a computerized player for the ABC TV game show "*Who Wants To Be A Millionaire?*". The player consists of a question-answering module and a decision-making module. The question-answering module utilizes question transformation techniques, natural language parsing, multiple information retrieval algorithms, and multiple search engines; results are combined in the spirit of ensemble learning using an adaptive weighting scheme. Empirically, the system correctly answers about 75% of questions from the *Millionaire CD-ROM, 3rd edition*—general-interest trivia questions often about popular culture and common knowledge. The decision-making module chooses from allowable actions in the game in order to maximize expected risk-adjusted winnings, where the estimated probability of answering correctly is a function of past performance and confidence in correctly answering the current question. When given a six question head start (i.e., when starting from the $2,000 level), we find that the system performs about as well on average as humans starting at the beginning. Our system demonstrates the potential of simple but well-chosen techniques for mining answers from unstructured information such as the web.

## 1 INTRODUCTION

Machine competence in games has long served as a benchmark for progress in artificial intelligence (AI). While we seem hardly close to building systems capable of passing a full-blown Turing Test, machine excellence in a growing number of games signals incremental progress. Games such as chess [13], checkers [27], Othello [7, 18], and Go

---

*This work conducted at NEC Laboratories America, Princeton, NJ.

[5] are formal enough to be solvable in principle, though are far from trivial to master in practice due to exponential size search spaces. In chess, checkers, and backgammon, current machine players rival their best human competitors. Recently, attention has turned to less structured game environments, like crossword puzzles [16], video games [30], and soccer [29], where game states, actions, or both are not easily enumerable, making a pure search formulation unnatural or impractical.

"*Who Wants to be a Millionaire?*" is a trivia game where actions are enumerable, though competence depends on the ability to answer general-interest questions—often requiring common sense or knowledge of popular culture—and to make decisions based on confidence, expected reward, and risk attitude. True human-level competence at *Millionaire* will likely require excellence in natural language processing and common sense reasoning. We present a first-order system that exploits the breadth and redundancy of information available on the World Wide Web to answer questions and estimate confidence, and utilizes a decision-theoretic subsystem to choose actions to maximize expected risk-adjusted payoffs.

## 2 RELATED WORK

### 2.1 QUESTION ANSWERING

A large body of research exists on question answering. For example, see the Question-Answering Track [32] of the Text Retrieval Evaluation Conference (TREC). Systems in this track compete against each other to retrieve short (50 or 250 byte long) answers to a set of test questions.

Question-answering systems typically decompose the problem into two main steps: retrieving documents that may contain answers, and extracting answers from these documents. For the first part of the task, retrieving a set of promising documents from a collection, the systems in the TREC QA track submitted the original questions to various information retrieval systems [32].

A number of systems aim to extract answers from documents. For example, Abney et al. [1] describe a system in which documents returned by the SMART information retrieval system are processed to extract answers. Questions are classified into one of a set of known "question types" that identify the type of entity corresponding to the answer. Documents are tagged to recognize entities, and passages surrounding entities of the correct type for a given question are ranked using a set of heuristics. Two papers [3, 21] present systems that re-rank and post-process the results of regular information retrieval systems with the goal of returning the best passages. These systems use the general approach of retrieving documents or passages that are similar to the original question with variations of standard TF-IDF term weight schemes [25]. The most promising passages are chosen from the documents returned using heuristics and/or hand-crafted regular expressions.

Other systems modify queries in order to improve the chance of retrieving answers. Lawrence and Giles [17] introduced *Specific Expressive Forms*, where questions are transformed into specific phrases that may be contained in answers. For example, the question "what is x" may be transformed into phrases such as "x is" or "x refers to". Joho and Sanderson [15] use a set of hand-crafted query transformations in order to retrieve documents containing descriptive phrases of proper nouns. Agichtein et al. [2] describe a method for learning these transformations and apply their method to web search engines.

Clarke, Cormack, and Lynam [8] describe a system that exploits the redundancy present in their corpus by using the frequency of each candidate answer to "vote" for the answer most likely to be correct. This approach is similar to the base approach of our system.

Recent work has shown that the Web can effectively be used as a general knowledge database for question-answering [9, 11, 22] and other related tasks. Fallman [12] presents a spelling and grammar checking tool that uses the *Google* search engine as its source of information, allowing it to handle names as well as informal aspects of a language such as idioms and slang expressions.

In contrast to most previous research, where systems are designed to search for an unknown answer, we present a system that aims to select the correct answer from a number of possible answers.

## 2.2 DECISION MAKING

Decision theory formalizes optimal strategies for human decision making [23], justified on compelling axiomatic grounds [26, 31]. The likelihood of future states is encoded as a subjective probability distribution and the value of future state-action pairs is encoded as a utility function; the decision maker optimizes by choosing actions that maximize future expected utility. A growing subfield in AI employs decision theory as a framework for designing autonomous agents. When the agent's state space grows unmanageably large—as in many real-world settings—graphical models such as Bayesian networks [14] or influence diagrams [28] that can encode probabilities and utilities compactly are often used. In *Millionaire*, the space of possible outcomes is small enough that decision trees [23], that explicitly enumerate probabilities and utilities for all future possibilities, are sufficient.

## 2.3 GAME PLAYING

Board games have dominated much of the history of AI in game playing [5, 6, 13, 18, 27]. This paper follows instead in the tradition of the crossword-puzzle-solving program PROVERB [16]. Like PROVERB, our *Millionaire* player brings together technologies from several core areas of artificial intelligence (including information retrieval, natural language parsing, ensemble learning, and decision making) to solve a challenging problem that does not naturally conform to the game-tree method for solving board games. Other domains under recent and rapid investigation—that are also not easily amenable to tree enumeration—include video games such as *Quake* [30] and soccer [29].

The *Millionaire* game has been explored in some previous work. Vankov et al. presented an abstract decision-theoretic model of *Millionaire* that yields a strategy for a player to maximize expected utility; however, it is still left up to the player to actually answer questions and assess confidence.[1] Rump [24] uses *Millionaire* as an educational tool to present problems in decision analysis including probability estimation and calculating expected utility, problems that our system must address. Clarke et al. [8] apply their general-purpose question-answering system to a set of questions asked on the *Millionaire* TV show (naturally composed of more early-round questions), answering 76 out of 108 questions (70.4%) correctly.

## 3 PLAYING MILLIONAIRE

*Millionaire*, a game show on ABC TV in the United States, might be characterized as a cultural phenomenon, spawning catch phrases and even fashion trends. The show originated in the United Kingdom and has since been exported around the world. Computers are explicitly forbidden as contestants on the actual game show by the official rules, negating any dreams we had of showcasing our system alongside Regis on national TV. We wrote our player instead based on a home version of the game: the *Millionaire CD-ROM, 3rd edition*.

### 3.1 RULES OF THE GAME

In *Millionaire*, the player is asked a series of multiple-choice trivia questions. Each correct answer roughly doubles the current prize. An incorrect answer ends the game and reduces the prize to the amount associated with the last correctly-answered "milestone" question, or zero if no milestones have been met. Milestones occur at the $1,000 and $32,000 stages, after questions five and ten, respectively. Answering fifteen questions correctly wins the grand prize of one million dollars. The difficulty of the questions (for people) rises along with the dollar value.

---

[1]The paper describing the system, unfortunately, has been removed from the web.

At any stage, after seeing the next question, the player may decline to answer and end the game with the current prize total. Alternatively, the player may opt to use any or all available lifelines to obtain help answering the question. Players are allotted three lifelines per game. The three lifelines allow the player to (1) poll the audience, (2) eliminate two incorrect choices, or (3) telephone a friend.

Our system does not address some aspects of *Millionaire*. In particular, we do not attempt to play the *fastest finger* round that determines the next player from a pool of candidate contestants. Winning this round entails being the first one to provide the proper ordering of four things by the criteria given in the question (e.g., "Place these states in geographic order from East to West: Wyoming, Illinois, Texas, Florida."). To be competitive, an answer generally must be provided within several seconds. Our question-answering system is neither designed to answer questions of this nature nor is it capable of answering most questions quickly. We also do not address extraneous tasks that people must perform in order to play the game, including speech recognition, speech synthesis, motor skills, etc.

### 3.2 CHARACTERISTICS OF QUESTIONS

The *Millionaire* CD-ROM game contains 635 questions that are roughly comparable in nature and difficulty to those on the TV show. The game places the questions into seven difficulty levels. The lower difficulty levels contain more common sense and common knowledge questions, while the difficult questions tend to be much more obscure. Lifeline information is also provided in the game data and is used in our game model.

For exploring algorithms and tuning parameters, we used three random 90-question samples and one random 180-question sample. Various reports on these training samples are reported throughout Section 4. Final test results on all 635 questions are reported in Section 5.1.

### 3.3 OUR PLAYER

Our *Millionaire* player consists of two main components, a question-answering (QA) module for multiple-choice questions and a decision-making (DM) module. We describe each component in turn below.

## 4 THE QA MODULE

Our system exploits the redundancy present in text corpora to answer questions. More precisely, we use the idea that question words associated with the answer tend to appear and are more likely to be repeated in multiple documents that contain the answer. We use the World Wide Web as our data source and several search engines (most prominently *Google*) as our conduit to that data.

We bring together several AI techniques from information retrieval, natural language parsing, and ensemble machine learning, as well as some domain-specific heuristics, in order to select answers and generate confidence measures. This information is then fed into the decision-making module, described later, to actually play the game.

### 4.1 THE NAIVE APPROACH: COUNTING

Our basic approach was to query Google with the question along with each of the four answers. Google enforces a 10-term limit on searches, so we performed stopword filtering on the questions to shorten our queries. Because answers were entirely comprised of stopwords in some cases, we did not filter them. The program generated queries in the format *answer filtered-question* to help ensure that the answer words fit in under Google's 10-term limit.

The response to the question was normally the answer that produced the highest number of search results. However, a number of questions are "inverted" in the sense that the answer is the one that is unlike the other three. We are able to identify nearly all of these by the presence of the word "not" in the question. In such cases, we choose the answer yielding the fewest results. This baseline strategy answers about half of the questions correctly.

#### 4.1.1 Simple Query Modifications

To improve on this strategy, we empirically found a small number of query transformations and modifications that increased the percentage of correct responses to 60%.

- Multiple-word answers are enclosed in quotes to require that they appear as a phrase in any search results.

- "Complete a saying" questions, identified by the presence of one of the strings "According", "said to", or "asked to", were handled by constructing each possible saying from the choices and requiring that it appear in the search results.

- When a query returns no results for any of the answers, we use a series of "fallback" queries that progressively relax the query. Quotes and words were removed from each query until at least one answer produced a non-zero number of search results.

- Longer web pages tend to contain lists of links, essays, manifestos, and stories; in general, their content is less useful for answering questions. Since search engines typically do not provide query syntax for restrictions on page size, we used a first-order approximation where we excluded .pdf files from the results.

Table 1: Pseudocode for DistanceScore, our proximity scoring method for favoring question words that appear near (within *rad* words of) answer words.

```
// wordList is the document split at spaces
DistanceScore(wordList, qWords, aWords, rad)
  score, answerWords = 0
  for i = 1 to |wordList| do
    if wordList[i] is in aWords then
      answerWords = answerWords + 1
      for j = (i-rad) to (i+rad) do
        if wordList[j] is in qWords then
          score += (rad - abs(i-j)) / rad
  if answerWords == 0 then return 0
  else return score / answerWords
```

## 4.2 WORD PROXIMITY MEASURES

Our heuristics for finding phrases are a specific variation of the general strategy of using proximity. Our belief—and that of many of the teams working on the TREC question-answering track [32]—is that not only do answers appear in the same documents as questions, but that they usually appear near the question words. In order to test proximity measures, we downloaded the first 10 (or all, if there were less than 10) pages Google returned for each query. We score each document based on a heuristic named *DistanceScore* that gives more credit to question words that appear closer to answer words in the document. Each such question word contributes a score between 0 and 1 to the score depending on how close the word is. A *radius* parameter controls what is considered near and how much a word adds to the score. We use the average score per answer word in the document to further penalize documents where answer words appear frequently but question words do not. Table 1 gives pseudo-code for DistanceScore.

Figure 1 shows the performance DistanceScore at various values for the radius on three 90-question samples, along with the performance of the naive method. Small random question samples were used to reduce the download and computation time required. DistanceScore performs reasonably well, doing worse than the naive method at low radius values but overtaking it at higher ones.

### 4.2.1 A Third Expert: Noun-Phrase Proximity

We developed a third strategy, also based on proximity. Since requiring multi-word answers to appear as phrases in web pages improved the accuracy of the naive method, another plausible strategy is to do the same for each of the noun phrases contained in the question. Noun phrases were identified using simple heuristics based on Brill's Part-of-Speech tagger. We submitted each {noun-phrase, answer} pair to Google and scored the results the same way as before. The result-count method produced poor results; however, downloading the returned documents and using DistanceScore to score each document worked well and produced results comparable to the previous two strategies.

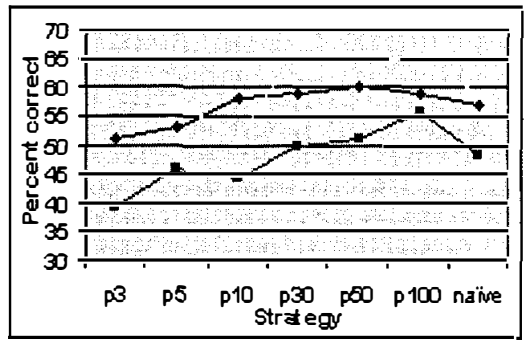

Figure 1: Question-answering accuracy versus proximity radius when using DistanceScore, as compared to the naive method on three 90-question samples. Each line represents performance on one sample.

### 4.3 COMBINING STRATEGIES

Among the naive, DistanceScore based on naive, and DistanceScore based on noun phrase strategies, at least one has the correct answer for about 85% of the questions in a 180-question sample. To exploit this, we look to answer-combining ("ensemble") approaches used commonly in machine learning, as summarized in [10].

Using the following formula, we attempt to combine our three strategies, or "experts," and produce a single score for each possible answer:

$$c_i = \sum w_S * (S_i/max\{S_{1..n}\}) \text{ over all strategies}$$

where $c_i$ is the combined score for answer $i$, $w_S$ is the weight for strategy $S$, $S_i$ is the score for strategy $S$ for answer $i$, and $n$ is the number of candidate answers.

Using the above formula to score candidate answers, we were able to reach 70% performance on the question sample. The weights yielding this, found empirically, were around ±0.05 of $w_n = 0.40$, $w_p = 0.15$, $w_{pp} = 0.45$ for the naive, word proximity, and noun phrase proximity strategies, respectively.

### 4.3.1 Combining Search Engines

In addition to combining strategies, we investigated using multiple search engines to improve results. We modified each of the three strategies to submit queries to AllTheWeb, MSN Search, and AltaVista, using syntax appropriate for each engine. The scores are combined using the same formula as above. Table 2 shows the results for each strategy using each search engine.

Google performs better than the other engines individually. However, we can combine the results from multiple engines, much as we combined the opinions of multiple strategies. Manually choosing a single set of weights for each {method, engine} pair showed that combining results across engines could result in better performance. For example, combining Google with AltaVista results in 75% of the 180-question sample being answered correctly.

Table 2: Performance of the three strategies using different search engines.

| engine | naive | proxim | phr prox | combined |
|---|---|---|---|---|
| Google | 55.6% | 55.0% | 68.9% | 70% |
| AllTheWeb | 56.1% | 51.7% | 58.3% | 66% |
| MSN | 44.4% | 48.9% | 47.2% | 58% |
| AltaVista | 46.7% | 55.6% | 56.1% | 68% |

#### 4.3.2 Confidence-Based Weight Assignments

However, choosing the weights manually was difficult. The optimal weights are probably sample-dependent and prone to overfitting; minor changes often led to 2-4% drops in performance. We modified our formula to assign different weights to each scoring method on a question-by-question basis, using the "confidence" of each scoring strategy $S$:

- Let $x_S$ be the "confidence ratio" defined by

$$x_S = \begin{cases} lowestscore/secondlowestscore & \text{if "not" question;} \\ secondhighestscore/highestscore & \text{otherwise.} \end{cases}$$

- Let $T = \sum(1 - x_i^4)$ over all strategies

- The weight for strategy $S$ is $w_S = (1 - x_S^4)/T$.

This assigns higher weights to more confident scoring methods. We chose the ratio between the second-best and best scores because we found a large difference in the ratio when the correct answer has the best score (mean ratio of 0.34) versus when the incorrect answer has the best score (mean of 0.58). Using these confidence-based weights generally results in slightly worse performance than hand-tuned weights, with Google falling to 69%, AltaVista to 65%, and the combination falling to 74% on the 180-question sample. Nonetheless, we believe that automatic confidence-based weights are more robust and less prone to overfitting than hand-tuned weights.

## 5 DISCUSSION: QA MODULE

Below we discuss several issues that came up in the course of building the question answering subsystem, and ways in that it could be improved.

### 5.1 OVERALL PERFORMANCE

We used confidence-based weights with the three-strategy method on the entire set of 635 *Millionaire* questions. The Google-based question-answerer got 72.3% of the questions correct, while one that used Google and AltaVista got 76.4%. On a set of 50 non-*Millionaire* trivia questions obtained from the shareware trivia game "AZ Trivia," the Google and AltaVista-based answerer answered 72% of the questions correctly.

We consider this to be good performance over the unstructured (and not necessarily correct!) data available from the web, supporting our claim that the web can be an effective knowledge base for multiple-choice question-answering.

### 5.2 CHOOSING GOOD WEIGHTS

In a few cases, using confidence scores to combine strategies caused the system's accuracy to fall below that of the best single strategy. This probably means that the "confidence ratio" is not a good heuristic for all scoring methods. The ratio is also difficult to compare between different engine-method combinations. For example, All The Web's ratios with the proximity score are consistently low, which translates into high confidence for many questions—even though this strategy only answers about half the questions correctly. Conversely, Google's ratios with the noun-phrase proximity score (which performs excellently) are consistently high, leading to lower confidences. The PROVERB crossword puzzle solver [16], which utilizes a similar approach to consider candidate answers from multiple experts, avoids this problem by allowing each expert to supply its own estimated confidence explicitly rather than applying a single function to every expert.

### 5.3 SAMPLE "PROBLEM" QUESTIONS

It appears that we have run into another example of the 80-20 rule. About one-quarter of the *Millionaire* questions are "hard" for the program. Below are examples of such questions that suggest areas in which a program trying to use the web as a knowledge base would need to improve.

**Common Sense.** *How many legs does a fish have? 0, 1, 2, or 4?* This information may exist on the web, but is probably not spelled out.

**Multiple Plausible Answers.** *What does the letter "I" stand for in the computer company name "IBM"? Information, International, Industrial, or Infrastructure?* "Information" probably appears just as often as "international" in the context of IBM.

**Polysemy.** *Which of these parts of a house shares its name with a viewing area on a computer screen? Wall, Roof, Window, or Basement?* The words "root" and "computer" often co-occur (e.g., the Unix superuser). This question also suggests that biases in the content of the web—originally by and for technical, computer-literate users—may hamper using the web as a general knowledge base in some instances.

**Non-Textual Knowledge.** *Which of these cities is located in Russia? Kiev, Minsk, Odessa, or Omsk?* The program doesn't know how to read maps.

**Alternative Representations.** *Who is Flash Gordon's archenemy? Doctor Octopus, Sinestro, Ming the Merciless, or Lex Luthor?* The word "archenemy" usually appears as two words ("arch enemy") on Flash Gordon (and other) pages.

## 6 THE DM MODULE

Answering questions is only half the battle. In order to actually play *Millionaire*, the system must also decide when to use a lifeline and when to "walk away". In order to compute its best next move, the decision-making module constructs a decision tree [23] that encodes the probabilities and utilities at every possible future state of the game. The full tree consists of decision forks for choosing whether to answer the question, use a lifeline, or walk away, and chance forks to encode the uncertainty of answering the questions correctly. The best choice for the program is the action that maximizes expected utility.

Utility is not necessarily synonymous with winnings in dollars. For example, suppose a contestant is at the $500,000 level. Even if he or she believes that by answering the final question his or her chances are fifty-fifty of winning either $1 million or $32,000 (expected value $516,000), the contestant will almost surely walk away with a guaranteed $500,000 instead. To model such *risk-aversion* we give the agent an exponential utility function $u(x) = 1 - e^{-(x/k)}$. For any finite $k > 0$, the agent exhibits risk averse behavior, though as $k \to \infty$, the agent becomes risk neutral (i.e., maximizes expected dollar value). In general, after playing many games, more risk averse agents will earn less prize money on average, though will have a smaller variation (standard deviation) of winnings.

### 6.1 MODELING THE GAME

We use the following specifications to construct the decision tree and play the game:

- For all questions beyond the current question, chance nodes are assigned probabilities based on historical past performance on a sample of questions from the associated difficulty level.

- For the current question (i.e., after the question has been asked and analyzed), the current chance node probability is $1 - x^\alpha$, where $x$ is the ratio between the second-highest score and the highest score obtained from the question-answering module (or the lowest score and the second-lowest score for "not" questions), and $\alpha$ is a tunable parameter that will be examined later. This lets us estimate confidence in our answer to the specific question being asked.

- The estimated future effect of lifelines on probability $p$ is given by the function $f(p) = -p^2 + 2p$, or the lifeline's performance based on historical data,

Table 3: Results of playing 10,000 games with $k = 250,000$ and $\alpha = 4$. The columns show the current prize level, number of games ending, number of correctly-answered questions, number of incorrectly-answered questions, times the player "walked away", number of lifelines used, number of lifelines that caused the the player to change its answer to the correct one, and number of lifelines that misled the player.

```
Stage   #-win  #-wrong #-right #-stop  llused  llgood  llbad
    0    4676      820    9180       0    1838     614      0
  100       1      781    8398       1    1653     535      0
  200       3      749    7646       3    1464     504      0
  300       5     1227    6414       5    2526     722     76
  500       7     1099    5308       7    2212     538     67
 1000    3700     1048    4260       0    2404     517     51
 2000      42      881    3337      42    1597     335     38
 4000      46      710    2581      46    1030     219     19
 8000      97      610    1874      97     625      62     21
16000      76      451    1347      76     388      48     10
32000     815      351     996       0     181      30     17
64000      37      254     705      37     118      11     11
125000     99      115     491      99     124      11      0
250000    156       56     279     156      72       1      1
500000    125       39     115     125      15       0      0
1000000   115        0       0     115       0       0      0
Avg. right: 5.29, winnings: $26328.87
```

whichever is greater. This models the idea that using a lifeline should raise the estimated probability of getting the question correct.

- When lifelines are used by the player, a new response and confidence level are calculated based on the new information received. For the 50/50 lifeline, the new response is simply the remaining choice with the higher score. The phone-a-friend and poll-the-audience lifelines are taken as an additional "expert" with a weight based on historical data.

### 6.2 PLAYING THE GAME: RESULTS

Table 3 shows the results of a risk-averse player ($k = 250,000$) playing 10,000 games using the above model using the question-answerer that uses Google and AltaVista. Questions were selected randomly from all the available questions in the appropriate difficulty level for each stage. Figure 2 summarizes the relationship between $k$, average winnings, and standard deviation. The more risk neutral the program is, the more it wins, and the more its winnings vary between games. Note that these points lie essentially along an efficient frontier (i.e., any gain in expected value necessitates an increase in risk [20]).

We also explored the effects of changes in $\alpha$, the exponent in the function used to convert confidence ratios into probabilities. Figure 3 graphs average winnings versus $\alpha$ and Figure 4 graphs the average number of correctly answered questions versus $\alpha$. Using higher $\alpha$ raises the program's estimated probability of answering a question correctly. Choosing $\alpha$ too low or too high hinders game performance since the program chooses to stop too soon or incorrectly answers questions that it is overconfident about. While high $\alpha$ values can produce high average winnings, it

Table 4: Human performance on the television show as reported on ABC's website in July 2001, compared to the computer's performance when given no handicap, and a six-question handicap.

```
Stage     human (pct)  computer (pct)  6-handi (pct)
     0     14  (2.0%)   4676 (46.8%)      0  (0.0%)
   100      0  (0.0%)      1  (0.0%)      0  (0.0%)
   200      0  (0.0%)      3  (0.0%)      0  (0.0%)
   300      0  (0.0%)      5  (0.1%)      0  (0.0%)
   500      0  (0.0%)      7  (0.1%)      0  (0.0%)
  1000    195 (28.6%)   3700 (37.0%)   5447 (54.5%)
  2000      0  (0.0%)     42  (0.4%)      0  (0.0%)
  4000      4  (0.6%)     46  (0.5%)     61  (0.6%)
  8000      9  (1.3%)     97  (1.0%)    249  (2.5%)
 16000     40  (5.9%)     76  (0.8%)    231  (2.3%)
 32000    166 (24.3%)    815  (8.2%)   2337 (23.4%)
 64000     92 (13.5%)     37  (0.4%)    139  (1.4%)
125000     89 (13.0%)     99  (1.0%)    370  (3.7%)
250000     48  (7.0%)    156  (1.6%)    504  (5.0%)
500000     18  (2.6%)    125  (1.3%)    311  (3.1%)
1000000     8  (1.2%)    115  (1.2%)    351  (3.5%)
Avg. winnings: $76497 vs. $26328.87 vs. $77380.90
```

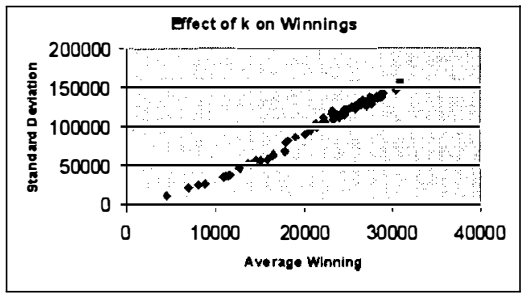

Figure 2: Standard deviation versus average winnings as $k$ ranges from 5,000 to 400,000 and $\alpha$ is fixed at 4. The gray point is a risk-neutral player. As $k$ increases, average winnings and standard deviation both increase.

comes at the cost of many more games (65%) resulting in a $0 prize as the player is too confident during early questions and saves its lifelines for later use. An $\alpha$ of 4 seems reasonable; about 47% of games result in $0 in that case, and the average winnings are relatively high.

## 7 DISCUSSION: DM MODULE

### 7.1 HARD QUESTIONS EASY, EASY ONES HARD

Table 4 compares the program's winnings to humans' winnings based on data from the ABC website as of mid-July, 2001. A striking feature of the program's performance is how often it wins nothing compared to people. Humans almost always answer the first several questions correctly; however, some are so obvious that the question-answerer cannot find the correct answer on the web. People generally do not encode common knowledge into their web documents. As a result, while the web seems to be a good knowledge repository for general knowledge, it is more difficult to use it as a common-sense database.

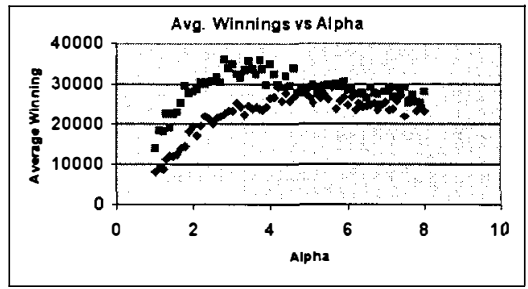

Figure 3: Average winnings versus $\alpha$. Black points are for a risk-averse player ($k = 250,000$); gray points are for a risk-neutral player.

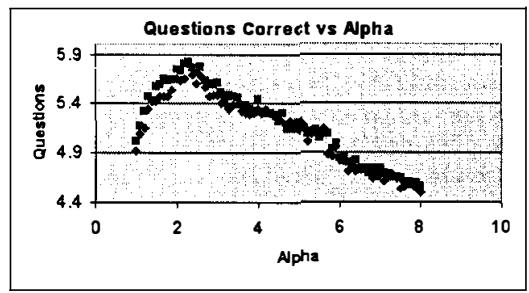

Figure 4: Average number of questions correct versus $\alpha$. Black points are for a risk-averse player ($k = 250,000$); gray points are for a risk-neutral player.

Observe that even if the question answerer could achieve a 95% success rate on early questions, it would still only have a 77% chance of achieving the $1,000 milestone. Its actual performance is worse, correctly answering 86% at level 1 ($100, $200, and $300) and 75% at level 2 ($500 and $1,000). Table 3 shows that as a result the program often exhausts its lifelines early in the game. On the other hand, we believe our program would have the upper hand against most people in a one-question, level 7, winner-takes-all match.

### 7.2 SIX QUESTIONS TO HUMAN

We might ask how well the program fares when given a handicap—that is, assuming that the program is able to answer the first $N$ questions correctly without using any lifelines. Figure 5 graphs the program's winnings versus its handicap. With a six question head start (going for $4,000) and all lifelines remaining, a risk-averse computer player ($k = 250000$) averages $77,381 with a standard deviation of $202,296.

Data from ABC's website as of mid-July, 2001 indicates that people on the show won about $76,497 on average with a standard deviation of $140,441. This suggests that, given a six-question handicap, the program performs about as well as qualified human players (i.e., those who self-selected to play the game, passed stringent entrance tests, and likely practiced for the game). Table 4 shows that even with the handicap, the program's performance is more variable than a human's, both winning big and losing early more often than people. However, its performance is still comparable, with "only" six "easy" questions separating the program from human-level performance.

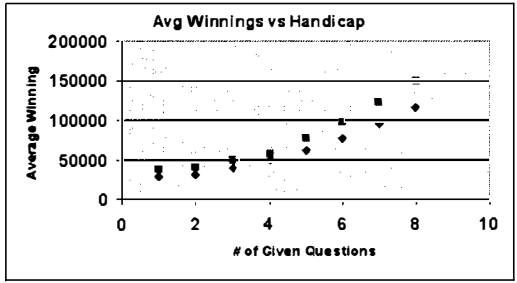

Figure 5: Average winnings versus handicap. Black points are for a risk-averse player ($k = 250,000$); gray points are for a risk-neutral player.

## 8 OTHER APPLICATIONS

While designed to play *Millionaire*, our system has other, more practical applications. The most straightforward is simply as a general-purpose question-answering system that can answer questions, provided a small pool of candidate answers can be provided or generated by some other means.

Combining the question-answerer with the decision maker can be useful in domains where a non-trivial penalty exists for answering a question incorrectly. For example, our system could be adapted to take the Scholastic Aptitude Test (SAT), an exam where answering a question incorrectly results in a lower score than not answering.

The general strategy of using search engines to mine the web as a giant text corpus shows promise in a number of areas. For example, web sites which provide content in multiple languages could become a knowledge base for automatic translation. Natural language processing programs could use the web as a corpus to help disambiguate parsing, or to find commonly occurring close matches to ungrammatical sentences.

## 9 CONCLUSIONS AND FUTURE WORK

We find that the web is effective as a knowledge base for answering generic multiple-choice questions. Naive methods that simply count search engine results do surprisingly well; more sophisticated methods that employ simple query modifications, identify noun phrases, measure proximity between question and answer words, and combine results from multiple engines do even better, attaining about 75% accuracy on *Millionaire* questions. When coupled with a decision-making module and given a six question handicap, our system plays the game about as well as people.

We believe that our system can be marginally improved in a variety of ways: for example, by employing better schemes for weighting multiple scoring methods, or by narrowing down the domain of a question and using domain-specific search strategies. We are also excited about the potential promised by approaches for structuring web data [4], although we believe that advances in automatic techniques for applying such structure (e.g., better natural language processing and common sense reasoning [19]) will be required for these approaches to succeed.

The call for such advances is a familiar one. From natural language processing to computer vision, a similar barrier exists across many subfields of AI: easy tasks (for people) are hard and hard tasks easy. While statistical and brute-force methods can go a long way toward matching human performance, an often difficult-to-bridge gap remains.